# Optimization in Software Engineering - A Pragmatic Approach

Guenther Ruhe

**Abstract**

Empirical software engineering is concerned with the design and analysis of empirical studies that include software products, processes, and resources. Optimization is a form of data analytics in support of human decision-making. Optimization methods are aimed to find best decision alternatives. Empirical studies serve both as a model and as data input for optimization. In addition, the complexity of the models used for optimization trigger further studies on explaining and validating the results in real world scenarios.

The goal of this chapter is to give an overview of the *as-is* and of the *to-be* usage of optimization in software engineering. The emphasis is on a pragmatic use of optimization, and not so much on describing the most recent algorithmic innovations and tool developments. The usage of optimization covers a wide range of questions from different types of software engineering problems along the whole life-cycle. To facilitate its more comprehensive and more effective usage, a checklist for a guided process is described. The chapter uses a running example *Asymmetric Release Planning* to illustrate the whole process. A Return-on-Investment analysis is proposed as part of the problem scoping. This helps to decide on the depth and breadth of analysis in relation to the effort needed to run the analysis and the projected value of the solution.

## 1 Introduction

The famous Aristotle (384 to 322 BC) is widely attributed with saying *It is the mark of an educated mind to rest satisfied with the degree of precision which the nature of the subject admits and not to seek exactness where only an approximation is possible* [43]. Applying Aristotle's saying to software engineering means that we need to understand the problem first, its nature and degree of uncertainty, before we start running any sophisticated solution algorithm. In this chapter, we explore the usage and usefulness of optimization techniques in software engineering. We take a pragmatic perspective and propose a model looking at the Return-on-Investment and provide directions for future research in this field.

The purpose of optimization is insight rather than numbers [26]. What counts is utilizing insight for making decisions. Software development and evolution are full of decisions to be made on processes, resources, artefacts, tools and techniques occurring at the different stages of the life-cycle. Some of these decisions have a strong impact on the success of the project. However, the information available for doing that is typically incomplete, imprecise, and even contradictory.

The paradigm of software engineering decision support [53] emphasizes the decision-centric nature of software engineering. It outlines how different methods including modelling, measurement, simulation as well as analysis and reasoning, can be used to support human decision-making. Optimization is another piece of that decision support agenda. Looking for the best possible item, artefact, process, action, or plan is tempting and a natural desire, but it is not automatically clear what that actually means. Part of the difficulty is that decisions typically are in the space of multiple criteria: Improving in one direction typically requires compromising against another criteria. In other cases, the data and information available are vague or even contradictory. Finally: To what extent decisions are based on rationality? Software development is a creative and human-centered process, and humans do not necessarily act based on rational arguments. So, another question arises: How valuable and how practical is optimization for the area of (Empirical) Software Engineering?

The chapter shows all the many opportunities of optimization in the various areas of software engineering (Section 9.2) and how to avoid pitfalls in its application. Optimization is not something that creates value automatically and easily. It is data intensive and requires time and effort investment. Optimization includes the whole process starting from the problem analysis, followed by modeling, running solution algorithms, understanding and interpreting data, likely modifying data and/or the underlying model, and re-running algorithms. In this chapter, we take a process view and provide a checklist for how to perform this process (Section 9.3). Its implementation is illustrated by a case study addressing asymmetric release planning (Section 9.4). Usage and usefulness of optimization is the key concern of Section 9.5. The chapter is completed by recommendations for further reading (Section 9.6) and conclusions on the future usage of optimization in the context of software engineering (Section 9.7).

## 2 Optimization in Software Engineering - Where?

The Capability Maturity Model Integration (CMMI) [13] characterized companies with the highest maturity CMMI level as *Optimizing*. This highest level of maturity refers to the application error analysis and process monitoring in order to optimize the current processes. While optimization is applicable to all types of questions, in software engineering it is primarily related to structured and semi-structured decisions on all technical and managerial aspects of software development and evolution.

The Software Engineering Body of Knowledge SWEBOK [1] lists technical and managerial areas to describe the field of software engineering:

- Software requirements
- Software design
- Software construction
- Software testing
- Software maintenance
- Software configuration management
- Software project management [1]
- Software engineering process
- Software engineering models and methods
- Software quality

Decisions are made during all stages of the software life-cycle. Depending on their type of control and impact, decisions are classified into operational (short-term), tactical (mid-term), and strategic (long-term) decisions [4]. As another dimension of their classification, decisions are classified, based on their input, into structured, semi-structured, or unstructured decisions. The emphasis of optimization is on tactical and strategic decisions based on structured or semi-structured information. In Table 1 (see Appendix) we present a list of publications using optimization methods in software engineering. As it is impossible to present a complete list, we only selected papers published since 2000. Among the many papers found, we picked those having highest number of citations in Google scholar (as of September 4, 2019).

[1] Replacing *Software Engineering Management* as used in SWEBOK

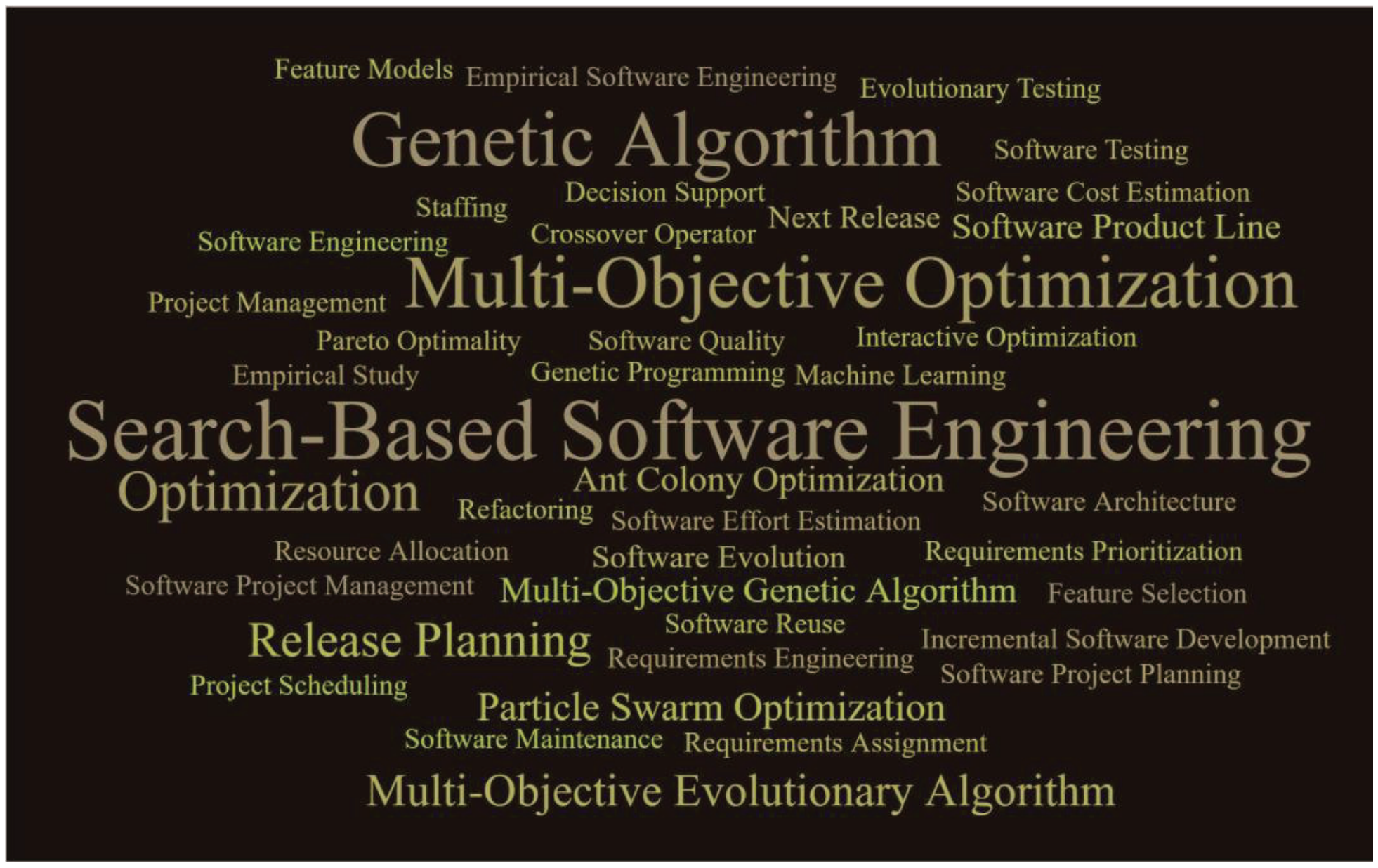

Figure 1: Word cloud for publications devoted to optimization in software engineering since 2000

Using the keywords of 84 optimization related papers published since 2000, a word cloud was created and is presented in Figure 1. We found a diversity of algorithms and concepts. Overall, the highest number of publications is in testing, requirements engineering, and design. This does not imply that there is nothing to optimize in the remaining areas. Lack of proper data might be one reason, and uncertainty in formulating explicit objectives and constraints another one for these current deficits.

# 3 Recommended Process and Checklist for Applying Optimization

## 3.1 Recommended Process

Optimization is a form of prescriptive analytics, aimed to propose actions to the decision maker. Different models for the process of data analytics (or data mining) exist, see the survey of Kurgan and Musilek [38]. We adapted the widely accepted CRISP data mining process introduced by Shearer [59] and integrated ideas of the engineering and the empirical cycle described by Wieringa [68]. The process model shown in Figure 2 establishes a link between (i) iterative software development (object of study), (ii) empirical studies (as a means to improve problem understanding, to create valid data, or to validate the research design), and (iii) existing model and data repositories.

For iterative development, two sample iterations $k$ and $k+1$ form the optimization context. Each iteration is assumed of having design, coding, and test activities, which result in a release version of the software product. Different parts of these processes might be optimized for higher efficiency. For example, there might be the problem of how to perform re-factoring, to finding best test strategies and test cases, to perform scheduling and staffing, or to decide about the functionality of the upcoming releases. For any of these questions, optimization helps to find a good or even the (formally) best answer. The process to find these answers is composed of eight steps that are further outlined in Section 3.2 by providing checklist questions to each step.

## 3.2 Checklist

Checklists are proven to be a means to facilitate the proper application of the overwhelming amount of existing knowledge existing in a field [25]. Similar to tools, applying checklists are no guarantee for success, but hopefully serve as filters or recommendations. They need continuous adaption to accommodate all the new directions and developments.

Checklists have been used in software engineering, e.g., for case studies, project risk analysis, and performing inspections. Here, we propose a checklist for the usage of optimization in software engineering. The checklist questions follow the key steps recommended in Section 3.1. The questions are classified in terms of their usage. M(andatory) questions need to be answered with ”yes” to make optimization a valuable effort. C(larification) questions (What?) are aimed to qualify the setup of the whole process. TMTB questions (How? How much?) are the ones that benefit from ”the more the better”.

**1 Scoping and ROI analysis:** Scoping defines the problem context and its boundaries. At this stage, performing an analysis of the potential *Return-on-Investment (ROI)* helps to determine the depth and the breadth of the investigation.

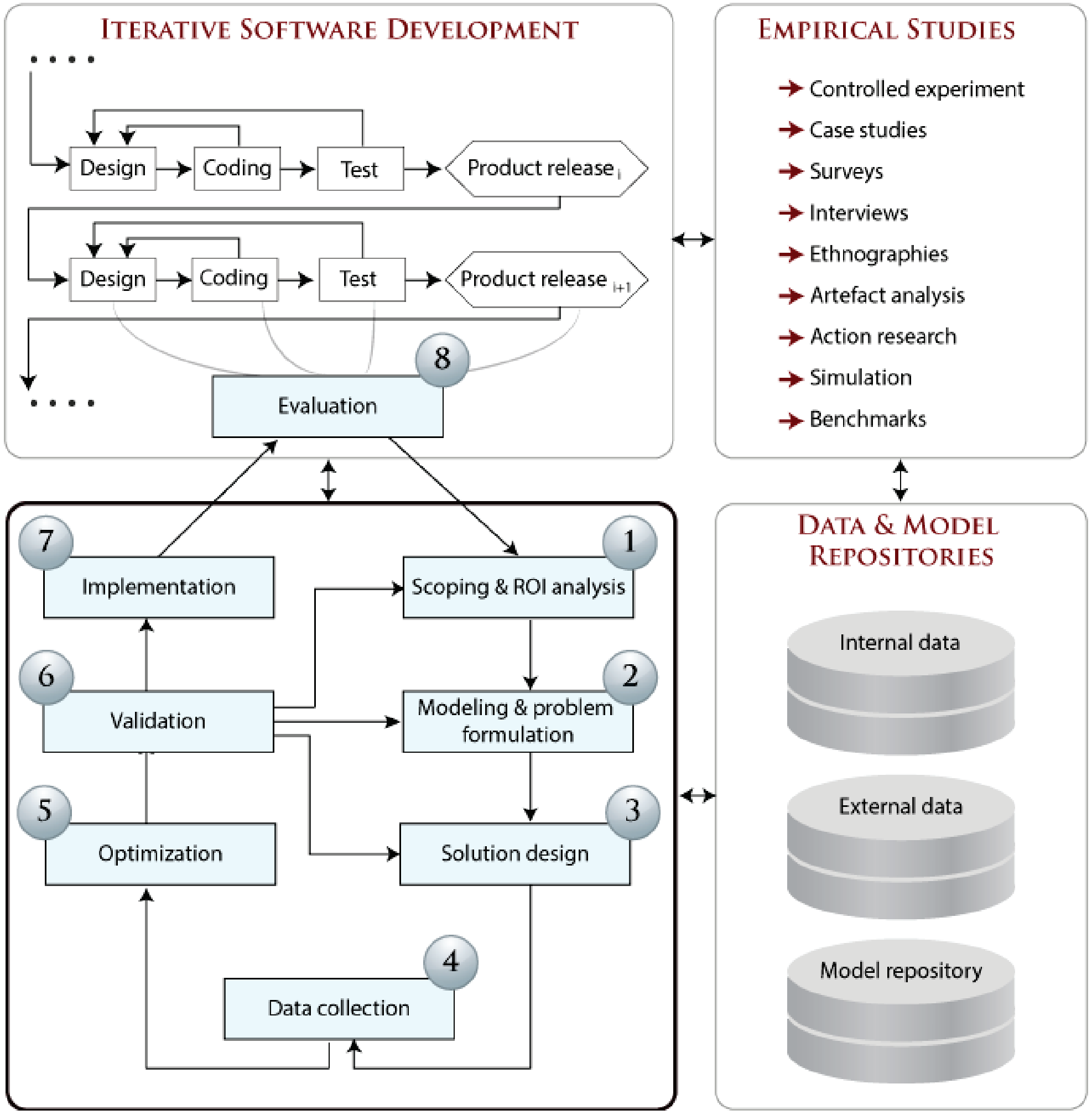

Figure 2: Process of performing optimization in the context of empirical studies for iterative software development

| Questions | Type |
|---|---|
| How important is the problem? | TMTB |
| How much time and money can be invested to solve the problem? | TMTB |
| Can the problem not be solved easily without optimization? | M |
| What are alternative solution approaches? | C |
| How much is the optimization problem aligned with business objectives? | TMTB |
| How much impact has an optimized solution in the problem context? | TMTB |
| What is part of the investigation, what is not? | C |
| Who are the key stakeholders and decision-makers? | C |

**2 Modeling and problem formulation:** Modeling and formulation of the problem is the phase where the variables, constraints, and objectives of the problem are formulated. The formulation needs to be verified against the original problem statement.

| Questions | Type |
|---|---|
| What are the key independent problem variables? | C |
| What is a reasonable granularity for problem formulation? | C |
| What are the dependent variables? | C |
| What are the human resource, budget, and time constraints? | C |
| What are the technological constraints? | C |

**3 Solution design:** The design step includes an analysis of what? and how much is enough? Exploring possible solution alternatives and its related tools depends on the scope selected. The design depends on the size of the problem (and its subsequent computational effort), the nature of the problem (linear, integer, convex, non-convex), and its projected impact. The design also decides between the use of traditional methods (linear, integer programming), and one of the many existing bio-inspired algorithms [7].

| Questions | Type |
|---|---|
| What baseline solution exist to compare with? | C |
| What are possible solution method alternatives and which ones have proven successful in similar context? | C |
| Which related tools exist? | TMTB |
| What are the expectations on optimization solutions (heuristic vs approximation vs exact)? | C |
| Is it search for just one (optimized) solution or for a set of solutions? | C |
| What scenarios are planned for running the algorithms? | C |
| How much would the optimization process benefit from interaction with the decision makers? | TMTB |

**4 Data collection:** Collect and prepare data needed to run optimization. While data is seldom complete, empirical studies can be used to improve the amount and quality of data. Goal oriented measurement [63] is the established technique to guide the data collection and analysis.

| Questions | Type |
|---|---|
| Is all necessary information available? | M |
| Is all available information also reliable? | TMTB |
| Is there a need for data cleaning? | M |
| Is there agreement between stakeholders on the data? | TMTB |

**5 Optimization:** Specify parameters for the tool and algorithm to execute optimization. How robust is the solution against changes on the input? What is the impact of adding or changing constraints?

| Questions | Type |
|---|---|
| Which parameter settings are made and why? | C |
| Should the parameter settings be varied and if so, how? | C |
| For randomized algorithms (e.g., bio-inspired algorithms), how many replications are needed to make sound conclusions? | C |
| Is there a time constraint for running the solution algorithm? | C |
| What are the termination criteria for the optimization algorithm? | C |

**6 Validation:** The solution of the mathematical optimization problem needs to be validated. How do the results compare to apply-ing alternative algorithms? What do the results mean for the original problem? Possibly, the scope, the problem formulation or the solution design need to be adjusted.

| **Questions** | **Type** |
|---|---|
| How much do the generated solutions make sense in the problem context? | TMTB |
| How much do stakeholders agree with the proposed solution(s)? | TMTB |
| In case of conflicts, how they can be resolved? | C |
| How robust is the proposed solution against changes of input data? | TMTB |

7 **Implementation:** The selected solution is implemented as a decision for the original problem.

| **Questions** | **Type** |
|---|---|
| Are the additional considerations to select one solution for final implementation in the original problem context? | C |
| Is there any need to adjust the proposed solution to the actual problem context? | C |

8 **Evaluation:** The usefulness of the implemented solution is evaluated in the original problem context and with the stakeholders involved.

| **Questions** | **Type** |
|---|---|
| How much does the implemented solution solve the original problem? | TMTB |
| How much the implemented solution is accepted by included stakeholders? | TMBM |
| How much the implemented solution improves the baseline? | TMTB |

The checklist serves a a filter. In the case that the two M questions are not answered positive, optimization is not recommended. The more information can be provide on C questions, the more guidance is available to run the process. For the TMTB questions, they more they accumulate ”more” evaluation, the more likely there is a positive return on the optimization investment.

# 4 Optimization Case Study: Asymmetric Release Planning

In this section, we go through a case study presented in [47]. The purpose is to illustrate the steps of the process illustrated in Figure 2.

## 4.1 Scoping and ROI Analysis

Release planning is a key part of iterative development. It is the process to decide about the functionality of upcoming releases of an evolving software product. Typically, a large number of requests for new or changing features as well as bug fixes are candidates to get implemented in each release. For simplicity, we call them features here. There are dependencies between the features that need to be considered for their implementation.

Implementation of features is expected to create value. However, there is an asymmetry in the sense that providing a feature does create satisfaction, but not providing it does not automatically create the same amount of dissatisfaction. This problem is called *Asymmetric Release Planning*. The ROI is supposed to be high, especially for products in a competitive market and when shipped in large quantities. The attractiveness of features is critical for success or failure of a product. Looking into the best release strategy in consideration of the asymmetry in value creation improves the traditional perspective of just looking at customer satisfaction as the single criterion.

## 4.2 Modeling and Problem Formulation

To run optimization, we need to model the problem and provide a formal problem description. Let $F = \{F(1) \ldots F(N)\}$ be a set of $N$ candidate features for development during the upcoming $K$ product releases. A feature is called *postponed* if it is not offered in one of the next $K$ releases. Each release plan is characterized by a vector $x$ with $N$ components $x(n)(n = 1 \ldots N)$ defined as:

$$x(n) = k \quad \text{if feature } F(n) \text{ is offered at release } k \tag{1}$$

$$x(n) = K + 1 \quad \text{if feature } F(n) \text{ is postponed} \tag{2}$$

The objective of the planning approach is to maximize stakeholder satisfaction and simultaneously minimize stakeholder dissatisfaction. These two objectives are independent and competing with each other. Pursuing each objective in isolation will create different release planning strategies.

To calculate stakeholder satisfaction $S(n)$ of feature F(n), all stakeholder responses related to satisfaction elements (Attractive and One-dimensional) are divided by the sum of the *Attractive*, *One-dimensional*, *Must-be* and *Indifferent* portions of that feature:

$$S(n) = \frac{F_A(n) + F_O(n)}{F_A(n) + F_O(n) + F_I(n) + F_M(n)} \qquad (3)$$

Similarly, $DS(n)$ is calculated by adding all responses with dissatisfaction elements (One-dimensional and Must-be) and dividing it by the total amount of relevant responses:

$$DS(n) = \frac{F_M(n) + F_O(n)}{F_A(n) + F_O(n) + F_I(n) + F_M(n)} \qquad (4)$$

For modeling of the total satisfaction objective, we follow the proven concepts of the EVOLVE based algorithms, in particular, the more recent EVOLVE II [54]. For a given time horizon of $K$ releases, there is a discount factor making the delivery of a feature less satisfactory when it is offered later. While using a weighting (discounting) factor $w(k)$ for all releases $k = 1...K$, we assume that $w(K + 1) = 0$, $w(1) = 1$ and

$$w(k) > w(k+1) \qquad (k = 1 \ldots K - 1) \qquad (5)$$

This assumption implies that the value of delivering a feature will be the higher the earlier it is delivered. For a plan $x$ assigning features to releases, *Total Satisfaction* $TS(x)$ is defined based on the summation of the discounted feature values $S(n)$ taken over all assigned features and all releases.

$$TS(x) = \sum_{k=1...K} \sum_{n:x(n)=k} w(k) \times S(n) \rightarrow Max! \qquad (6)$$

*Total Dissatisfaction* $TDS(x)$ of a plan $x$ follows the same idea as just introduced for satisfaction. The longer a feature is not offered, the higher the dissatisfaction. Similar to satisfaction, we introduce factors describing the relative degree of dissatisfaction between releases. $z(k)$ is the dissatisfaction discount factor related to release $k$. As dissatisfaction of non-delivery increases over releases, we assume $z(1) = 0$, $z(K + 1) = 1$ and

$$z(k) < z(k+1) \quad (k = 1 \ldots K) \qquad (7)$$

If plan $x$ would not offer any features at all, total dissatisfaction $TDS(x)$ would be the summation of all feature dissatisfaction values. More generally, if a feature is offered in release $k$, then this creates a dissatisfaction of $z(k) \times DS(n)$. If it is offered in the next release, no dissatisfaction is created at all. Total dissatisfaction TDS(x) created by a plan $x$ is modeled as the summation of all adjusted feature values $DS(n)$, and this function needs to be minimized:

$$TDS(x) = \sum_{k=1 \ldots K+1} \sum_{n:x(n)=k} z(k) \times DS(n) \rightarrow Min! \tag{8}$$

Implementation of features is effort consuming. We make the simplifying assumption of just looking at the total amount of (estimated) effort needed per feature. The estimated effort for implementation of feature $F(n)(n = 1 \ldots N)$ is denoted by *effort(n)*. When planning $K$ subsequent releases, the consumed effort per release is not allowed to exceed a given release capacity. For all releases $k$ $(k = 1 \ldots K)$, this capacity is denoted by $Cap(k)$.

More formally, a *feasible release plan x* needs to satisfy all constraints of the form:

$$\sum_{n:x(n)=k} effort(n) \leq Cap(k) \quad \text{for k} = 1 \ldots \text{K} \tag{9}$$

In Equation (9), effort consumption for release $k$ is constrained by the given capacity $Cap(k)$. For each release $k$, the summation is done over all the features $F(n)$ that are assigned to this release, i.e. fulfilling $x(n) = k$.

Among all the plans fulfilling resource constraints (known as *feasible plans*), a plan $x^*$ is called a *trade-off solution* if no other plan exists that is better on one criterion and at the same time not worse in the other. This means that we are looking for feasible plans $x^*$ with the property that there is no other feasible plan $x'$ (also called a *dominating plan*) such that is better in one dimension and not worse in all the others.

**Asymmetric Release Planning ARP:** We consider a given set of features $F(n)$ with feature values $S(n)$ and $DS(n)(n = 1 \ldots N)$. Among all the plans fulfilling resource constraints, the ARP problem is to find trade-off solutions for concurrently maximizing $TS(x)$ and minimizing $TDS(x)$. That means, ARP is the problem of finding trade-off release plans that are balancing satisfaction and dissatisfaction.

## 4.3 Solution Design

Software release planning can be solved by a variety of algorithms. For a more recent analysis we refer to [2]. In its simplest form, greedy heuristics could be applied. The general greedy principle is to select the best local features at each iteration, where the definition of locally best varies between the heuristics [15]. With no backtracking, greedy solutions are fast and often ”good enough”. The quality of the solutions depend on the problem structure and the instance of the problem. It can be quite far from the optimum in specific instances.

Integer linear programming was used by Veerapen et al. [64] to solve the single and bi-objective Next Release Problem. While there has been a dominance of search-based techniques in the past (starting with the genetic algorithm of Greer and Ruhe [27]), the authors have shown that integer linear programming-based out-performs the NSGA-II [17] genetic approach on large bi-objective instances. For the bi-objective asymmetric release problem, we propose a method of solving a sequence of single-criterion optimization problems, each of them generating a new or an existing trade-off solution. The step-size for varying the parameter can be selected by the concrete problem. For the implementation, we apply the (mixed) integer linear programming optimizer Gurobi[30] version 6.4 and its interface MATLAB to manage data.

## 4.4 Data Collection

Data collection covered towards elicitation of features, effort estimation, and feature evaluation by stakeholders. The lists of features, effort estimation, continuous Kano survey, and results are available online[2].

**Stakeholders:** We invited 24 software engineering graduate students to serve as stakeholders. Even though students were not a direct customer of the company, they were familiar with the domain and were considered to be representative for the purpose of this case study.

**Weight of stakeholders:** The survey participants provided a self-evaluation in terms of their familiarity with Over-the-Top (OTT) services and mobile applications. At the beginning of the survey, stakeholders stated their domain expertise on a Likert scale ranging from one to nine. We used this value as the weight of stakeholders for the planning process.

**Features**: The pool of candidate app features was extracted from the description of 261 apps, all of them providing media content over the internet without the involvement of an operator in the control or distribution of the OTT TV services. A commercial text analysis tool was used to retrieve 42 candidate features. Domain experts evaluated the meaningfulness of extracted features and eliminated the phrases which did not point into any OTT feature. Feature extraction itself

[2] http://www.maleknazn.com/tools-and-datasets

was managed by the case study company and resulted in 36 features further investigated.

**Feature value:** To predict the impact of offering versus missing features, we applied the Kano analysis [35]. We performed a survey with a continuous Kano design and asked the two types of questions (functional and dysfunctional). For each feature, each of the stakeholders expressed the percentages that the feature matches one of the five possible answers per question.

**Effort:** The effort for developing each feature was estimated by domain experts within the company. A product manager and two senior developers estimated the effort needed (in person hours) to develop each feature. They applied a triangular (three point) effort estimation to estimate the optimistic, pessimistic and most-likely effort amount needed to deliver a feature. The three estimates were combined using a weighted average.

**Capacity**: To show the impact of tight, medium, and more relaxed resource availability, we ran three concurrent scenarios with three varying release capacities Cap(1) of 112.7 (lower bound), 367.4 (most probable), and 625.5 (upper bound) person hours, respectively.

## 4.5 Optimization

We were running the optimization solver for three defined optimization scenarios in correspondence to three varying capacity levels. Each time, a set of alternative solutions is generated. In total, 14 structurally trade-off solutions (plans) were generated. Each plan represents one possible way to balance between satisfaction and dissatisfaction of stakeholders. Based on the equivalence between parametric and multi-objective optimization [14], each solution set is received from running a sequence of single-criterion problems.

## 4.6 Validation

We performed a comparison between the quality of the optimized solutions in comparison to those obtained from (i) random search and (ii) heuristic search. Using random search is selecting a feature randomly as long as the effort for implementing that feature is less than the available capacity. The results showed that optimized solutions strongly dominate all the 1,000 solutions generated by random searching.

We also compared the results against running eight different heuristics. 66.7% of the heuristic plans were dominated with at least one of the fully optimized solutions. This demonstrates that heuristics are fast and conceptually easy, but often not good enough. The compre-hensiveness of solutions generated from the optimization in conjunction with their guaranteed quality is considered a strong argument in favor of the proposed approach.

### 4.7 Implementation and Evaluation

We conducted a survey to understand stakeholders preference among the various plans generated. The survey included 20 stakeholders. Using Fleiss Kappa test [60] for measuring inter-rater agreement showed a slight to poor agreement between the 20 participants.

By comparing plans per stakeholder, among stakeholders and between criteria, we found that:

- One plan does not fit all: For both planning objectives, there is substantial variation between stakeholders in terms of what they consider their preferred solution.
- One criterion is not enough: Six of the 20 stakeholders have a varying top preference when comparing plans selected from satisfaction and dissatisfaction perspective.

## 5 Usage and Usefulness of Optimization

### 5.1 Limits of Optimization Models

As any model, even the most detailed models lack some details in comparison to the real-world ecosystem. This means that the representation of real world always lacks some details and contains inaccuracies. Meignan et al. [46] described four types of potential optimization model limitations:

- Approximation of complex problem's aspects: Formulation of real-world problems might be difficult because of the lack of quantifying constraints or objectives. Human involvement, as being the case in software engineering, often requires to use approximations of the phenomena.
- Simplification for model tractability: Even if the phenomena under investigation are quantifiable, simplification of the model is needed to apply a computational optimization approach. For example, linear models are often used for that purpose.
- Limited specifications: In some cases, the problem is not (yet) well defined. This might be caused by the lack of problem and domain knowledge.

- Lack of resources: Optimization is not for free but consumes time and cost. This implies making compromises and simplifications.

## 5.2 Difficulties which are Specific to the Discipline of Software Engineering

Software engineering differentiates from other disciplines by a number of factors [56]. Among those factors, some of them are critical for deciding and running optimization:

- High degree of uncertainty in the software project and product scope.
- Planning and estimating of software projects is challenging because these activities depend on requirements that are often imprecise or based on lacking information.
- Software development is non-deterministic. The data received from observations are incomplete and sometimes contradictory.
- Objective measurement and quantification of software quality is difficult.

As a conclusion of the above, we need to check more carefully which type of optimization is most appropriate and in which situation and how much we should pursue this pathway. Is it always ”The more the better”? The following subsection proposes three attributes to answer this question.

## 5.3 How Much is Enough?

There is a very broad spectrum of decision problems that potentially benefit from running optimization. But which technique to apply in what situation? Both the problems under investigation and the techniques available have characteristics that are relevant to decide which one is used for what. Along with this understanding goes the question *How much is enough?*. In the sequel, we define three key dimensions which have shown to have substantial impact on the breadth and depth of analytics of optimization. The key motivation for doing this is the statement attributed to Aristotle mentioned in the introduction.

### 5.3.1 Validity - How valid is the problem definition?

There is no easy way to measure validity of a model (being always wrong and sometimes useful), but some possibilities to verify its behaviour from running through some scenarios with an outcome already known. Validity can also relate to wickedness [52] which refers to the difficulty to find a proper formulation of the problem and a termination criterion for its solution.

#### 5.3.2 Cost - How much effort is needed to run the whole optimization process?

The cost of performing the process outlined in Figure 1 is mainly determined by the resources consumed in it. The cost might vary substantially between the cases. For example, the effort for data collection is highly influenced by the amount and quality of data already available and the effort needed to elicit additional information. Once the problem is defined, the notion of computational complexity [24] guides the effort estimation to determine a solution with a proven quality. This effort typically depends on the size (small, medium, or large number of variables involved), difficulty (low, medium, or high in terms of number of constraints, objectives), the linearity versus non-linearity, and the continuous versus discrete nature of the variables included.

#### 5.3.3 Value - How valuable is it to find an optimized solution?

How much value is added to the real-world problem when a optimized and carefully analyzed solution is applied versus an ad hoc one suggested by the domain expert? This value is associated with the impact of the decision to be made. Operational decisions are in the day-to-day business with typically short-term impact on a company. Tactical ones are concerned with work assignments, selection of tools, reuse of artefacts. Strategic decisions have the longest impact, but are also based on the strongest degree of uncertainty.

### 5.4 Return-on-Investment

Comparing the investment made into something with the potential return achieved from this investment is a common question, mainly triggered from economical considerations. For example, Erdogmus et al. [21] analyzed the ROI of quality investment. The authors state that ”We generally want to increase a software products quality because fixing existing software takes valuable time away from developing new software. But how much investment in software quality is desirable? When should we invest, and where?”

The same idea could be used to the application of optimization. More precisely, evaluating the impact of decisions made based on opti-mization versus the baseline decision-making approach, typically some form of relying on intuition. There are at least three prerequisites for doing that: (i) Characterization of the context, (ii) Characterization and quantification of the investment (cost), and (iii) Projection of the value added from implementing an optimized solution.

For the characterization of context, Dyba et al [20] introduced a template that has the two main dimensions called *omnibus context* and *discrete context*. The first dimension includes the 5 W's: What? Who? Where? When? and Why? The second one is along the technical, social, and environmental aspects of the context which are likely vary in detail from case to case. Specifying the context attributes allows to related specific empirical results to the characteristics where they are coming form. It also helps to provide more specific recommendations about when and how well certain things work.

Software development and evolution is a value-creating process. However, there seems to be a "disconnect" between the decision criteria that guide software engineers and the value creation criteria of an organization [8]. Projecting the impact of an improvement in decision-making against a technical parameter (e.g., test coverage) in terms of its impact can be expressed as *Net Present Value* (NPV) but is a challenging task. If analyzing a project over a period of n time periods, the formula for the net present value of a project is:

$$NPV = \sum_{t=0}^{n} \frac{R(t)}{(1+d)^t} \quad (10)$$

Therein, R(t) denotes the difference between net cash inflow and net cash outflow during a single period t. The model assumes a discount rate d to transfer future value to the today value. The added NPV that is from applying an optimized versus a baseline solution is the difference between their respective NPV's.

$$NPV_{Added} = NPV_{OptimizedSolution} - NPV_{Baseline} \quad (11)$$

The cost of analytics is determined from both the depth and the breadth of investigations. The *depth* can be exemplified by the extent of running optimization algorithms to achieve close to optimal results. Another example refers to the level of details included in visualization. The extent of applying multiple analytical techniques to data from multiple sources to achieve results refers to *breadth*. Solving a problem by looking for data from different sources and/or performing method triangulation is of clear value but also of clear additional cost. The variation of perspectives in visualization is an example for breadth.

Figure 3 shows a ROI curve of technology usage. Following some phase of increase, there is a saturation point. After that, further investment does not further pays off. We hypothesize a similar principal behaviour for the usage of optimization and analytics in general.

### 5.5 ROI of Asymmetric Release Plan Optimization

It is difficult to quantify the ROI. The value of providing the right features at the right time depends on the competitiveness of the market the product is related to, the number of product instances sold, and maturity of data collection and decision-making processes the organization is in. For the running example, additional effort was needed to elicit the information related to stakeholder satisfaction and dissatisfaction. We performed a survey with six company managers and asked: To what extent do you think the additional effort (from answering ten questions per feature based on Kano) is worthwhile? In response, five managers agreed or strongly agreed, and one manager was neutral about the efficiency of the Kano model.

## 6 Recommended Further Reading

Optimization is an established discipline with roots going back to names such as Leonid Kantorovich, George Dantzig, and John von Neumann. Its applications range through all disciplines of Science, Engineering, and Economics. There are numerous models, methods and tools. The encyclopedia of optimization [22] gives an overview. In its essence, optimization is the process of searching for the best out of a pool of alternatives. What means *best* is described by one or multiple criteria. The alternatives are explicitly or implicitly described (by constraints). For the actual optimization step in the whole problem solving process (Step 5 in Figure 1), different alternatives can be considered, ranging from simple heuristics to meta and hyper heuristics to exact common purpose solvers. The encyclopedia of optimization edited by Floudas  Pardalos [22] gives a good overview. In the sequel, we provide guidance for further reading for three emerging directions of using optimization in the context of (empirical) software engineering.

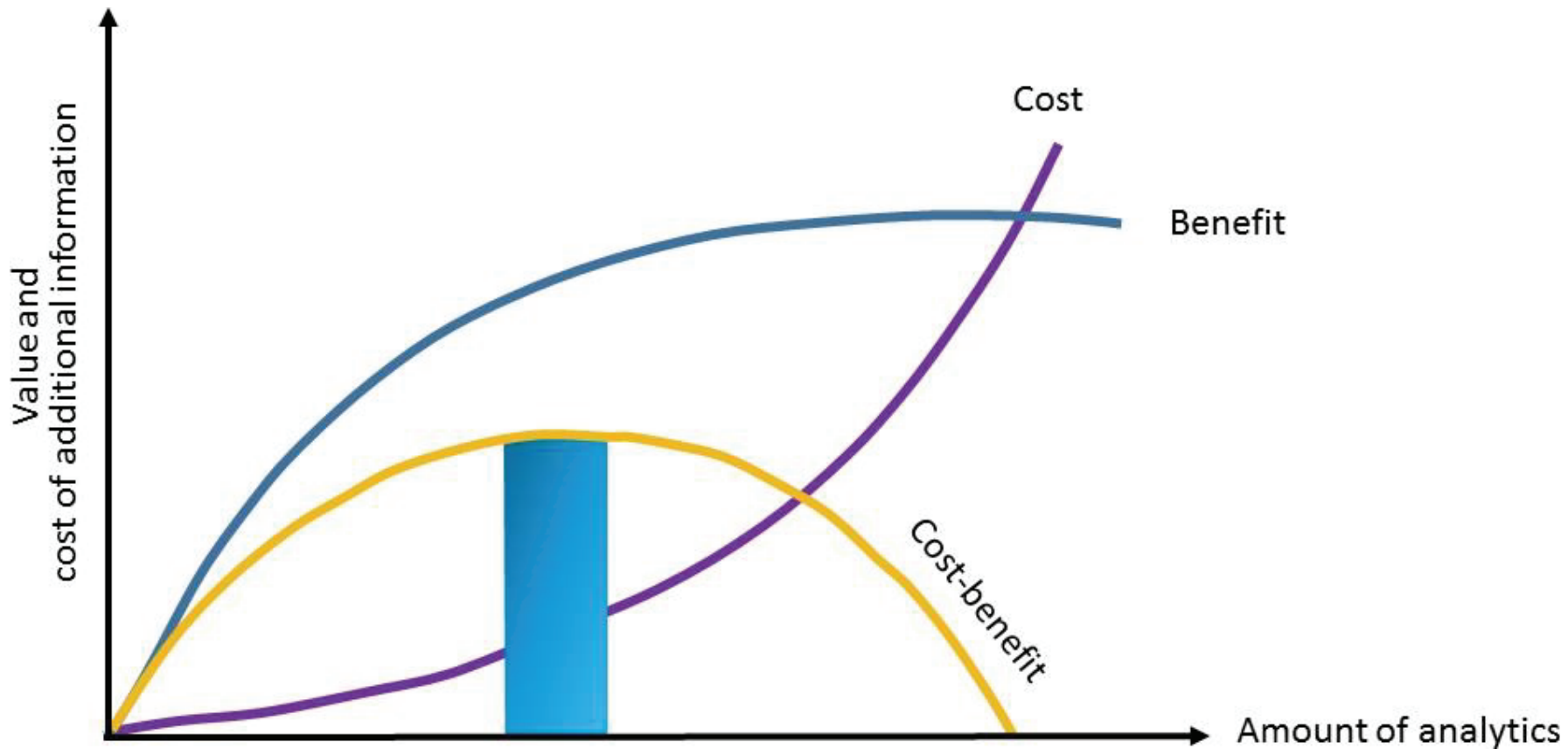

Figure 3: Expected ROI curve from technology investment.

## 6.1 Meta and Hyper Heuristics

In the area of software engineering, the term Search-based Software Engineering (SBSE) was coined by a paper of Harman and Jones [32], who argued that software engineering is ideal for the application of meta-heuristic search techniques, such as genetic algorithms, simulated annealing, and tabu search. In November 2017, the SBSE repository [71] at University College London counted 1727 relevant publications. More recently, the portfolio of techniques has been enlarged by other bio-inspired algorithms that are designed and following the behavior of biological systems. These algorithms are intuitive and have proven successful in many occasions including software engineering [36].

Independently, hyper-heuristics have been designed (see [10] for a state-of-the art survey). The key characteristic is that these algorithms operate on a search space of heuristics (or heuristic components) rather than directly on the search space of solutions to the underlying problem that is being addressed [10]. For the area of testing, Balera et al. [5] performed a systematic mapping study on hyper-heuristics. For the multi-objective next release problem and analyzing ten real-world data sets, Zhang et al. [73] have shown that hyper-heuristics are particularly effective.

## 6.2 Bio-inspired Algorithms

Bio-inspired optimization is an emerging paradigm which encompasses the principles and inspiration of the biological evolution of nature to develop new and robust optimization techniques [7]. These algorithms are drawing attention from the scientific community due to the increasing complexity of the problems, increasing range of potential solutions in multi-dimensional hyper-planes, dynamic nature of the problems and constraints, and challenges of incomplete, probabilistic and imperfect information for decision making. Not much exploration has happened in the true application space for the rest of these algorithms, probably due to the recency of some of these developments or due to the lack of availability of pseudo-codes which can be used directly. There is a need for studies highlighting the preferred application for algorithms like artificial bee colony algorithm, bacterial foraging algorithm, firefly algorithm, leaping frog algorithm, bat algorithm, flower pollination algorithm and artificial plant optimization algorithm in actual problem contexts.

## 6.3 Interactive Optimization

Interactive optimization approaches acknowledge existing limits to modeling and parameter settings and values the user's expertise in the application domain [46]. Interactive approaches maintain the human expert in the problem solving loop. Meignan et al. [46] distinguish between problem-oriented interaction and search-oriented interaction. For the first one, the user can either adjust or enrich the optimization problem, e.g., by adjusting existing constraints or objectives or by defining new ones. For search-oriented interaction, the user actively influences the search procedure, e.g., by parameter tuning.

# 7 Conclusions

Optimization is an important means to improve software development and evolution. Optimizing processes is the ultimate goal of most mature CMMI level 5 organizations. Practically, there is almost no limit to consider optimization in software engineering. But as for any technology, its usage needs to be guided. Besides the opportunity to enlarge the scope of optimization, this chapter emphasizes the need to take a closer look at the ROI of optimization. Taking into account the perceived validity of the problem formulation, the ROI model serves as justification for investing into optimality. In the language of the CMMI model, it is unlikely that a specific part of the software development landscape is highly optimized if the whole surrounding is immature. The ROI estimation servers as a guidance in this decision process of deciding How much optimization is needed and how much is enough?.

Optimization by no means is a silver bullet. We are not fascinated just by numbers, but are more interested in insight, especially actionable insight. If designed and performed properly (for example, as guided in the recommended process of Figure 1), it is a valuable part of decision support. Its pragmatic usage means to understand the scope and degree of its usage. Furthermore, it often means searching not for just one ultimate solution but for a diversity of alternative solutions which is formulated as the diversification principle [54]:

*A single optimal solution to a cognitive complex (optimization) problem is less likely to serve the original problem when compared to a portfolio of optimized solutions being qualified AND structurally diversified.*

# Acknowledgment

This research was supported by the Natural Sciences and Engineering Research Council of Canada, Discovery Grant RGPIN-2017-03948. The literature analysis of the study was supported by Debjyoti Mukherjee. The author is grateful to discussions with and comments received from Maleknaz Nayebi and Julian Harty.

# Appendix

| Abbreviation | Full Text |
|---|---|
| GA | Genetic Algorithm |
| SA | Simulated Annealing |
| NSGA - II | Non-dominated Sorting Genetic Algorithm II |
| MOEAs | Multi-Objective Evolutionary Algorithms |
| ACO | Ant Colony Optimization |
| ILP | Integer Linear Programming |
| QoS | Quality-of-Service |
| EAs | Evolutionary Algorithms |
| GSA | Genetic Simulated Annealing |
| SA/AAN | Simulated Annealing with Advanced Adaptive Neighborhood |
| CCEA | Cooperative Co-Evolutionary Algorithms |
| LSR | Least-squares Linear Regression |
| CBR | Case-based reasoning |

Table 1: Selected optimization studies in software engineering

| Area | Problem | Optim. Criteria | Optim. Method | Year | Citations[1] | Ref |
|---|---|---|---|---|---|---|
| Requirements | Selecting optimal features for next release | Value | GA | 2004 | 434 | [27] |
| Requirements | Next release problem | Cost, value | Pareto optimal genetic algorithm, NSGA - II | 2007 | 237 | [72] |
| Requirements | Software release planning | Maximize the value of the product for each release | Hybrid intelligence | 2004 | 125 | [55] |
| Requirements | Release planning for evolving systems | Value | GA | 2005 | 109 | [58] |
| Requirements | Software release planning | Selection of highly coupled features in the same release | ILP | 2007 | 103 | [57] |
| Requirements | Next release problem | Minimize cost and maximize customer satisfaction | GA, Evolutionary Strategy | 2011 | 85 | [19] |
| Requirements | Software release planning | Optimized resource allocation | ILP, GA | 2009 | 58 | [48] |
| Design | Modularization | Cohesiveness | Pareto optimal genetic algorithm | 2010 | 274 | [50] |
| Design | Refactoring | Search based refactoring | Pareto optimality | 2007 | 209 | [33] |
| Design | Performance of software | Genetic improvement | Genetic Programming | 2014 | 173 | [40] |
| Design | Modularization | Cohesiveness | GA | 2002 | 163 | [31] |
| Design | Feature selection | Value | GA | 2011 | 156 | [29] |
| Design | QoS-driven web service selection | Quick convergence | GA | 2008 | 156 | [44] |
| Design | Optimal use of the hardware | Performance | Profiling, optimization | 2003 | 134 | [37] |
| Design | Reassign methods and attributes to classes in a class diagram | Class coupling and cohesion | Multi-objective GA | 2010 | 107 | [9] |

[1] Google Scholar citations as of Sep. 5, 2019

Table 1: Selected optimization studies in software engineering

| Area | Problem | Optim. Criteria | Optim. Method | Year | Citations[1] | Ref |
|---|---|---|---|---|---|---|
| Design | Choosing the optimal architectural design alternative | Reduce development cost, improve quality | EAs | 2006 | 80 | [28] |
| Design, Maintenance | QoS-aware service composition | Concrete services combination | GA | 2008 | 98 | [65] |
| Testing | Regression test prioritization | Quality | GA | 2006 | 366 | [66] |
| Testing | Regression test case selection | Code coverage, past fault-history detection and execution cost | GA | 2007 | 333 | [70] |
| Testing | Test data generation | Multi-objective branch coverage | MOEAs | 2007 | 160 | [39] |
| Testing | Test case prioritization for regression testing | Fault coverage | ACO | 2010 | 80 | [61] |
| Testing | Comparing different algorithms for test case generation | Coverage | GA, SA, GSA, SA/AAN | 2007 | 76 | [69] |
| Project Management | Software effort estimate | Accurate effort estimation | GA with Grey Rational Analysis | 2008 | 238 | [34] |
| Project Management | Staffing of software project | Value | Constraint satisfaction | 2008 | 166 | [6] |
| Project Management | Software effort estimate | Input feature subset, parameters for machine learning | GA | 2010 | 139 | [49] |
| Project Management | Project task scheduling and human resource allocation | Flexibility | ACO | 2012 | 138 | [12] |

[1]Google Scholar citations as of Sep. 5, 2019

Table 1: Selected optimization studies in software engineering

| Area | Problem | Optim. Criteria | Optim. Method | Year | Citations[1] | Ref |
|---|---|---|---|---|---|---|
| Project Management | Comparing different techniques for planning resource allocation | Duration | GA, SA, Hill Climbing | 2005 | 128 | [3] |
| Project Management | Software project scheduling | Scheduling | GA | 2008 | 121 | [11] |
| Project Management | Assign features to releases | Value | ILP | 2008 | 108 | [62] |
| Project Management | Comparing software effort prediction techniques | Accuracy of prediction | Expert judgment, LSR, CBR | 2003 | 105 | [45] |
| Project Management | Software cost estimation | Feature selection with lower complexity | Mutual information based feature selection | 2009 | 95 | [41] |
| Project Management | Deciding whether to buy or build a component | Cost and quality optimization | ILP | 2008 | 93 | [16] |
| Project Management | Allocation of testing resource | Reliability and testing cost | MOEAs | 2010 | 87 | [67] |
| Project Management | Use of search-based optimization techniques for management activities | Staff and task allocation | GA, NSGA II | 2011 | 58 | [18] |
| Project Management | Software project staffing and job scheduling | Team staffing and work package scheduling | CCEA | 2011 | 51 | [51] |
| Process | Cloud computing deployment and re-configuration | Response time and cost | GA | 2013 | 125 | [23] |
| Quality | Software quality modeling | Cost | Genetic Programming | 2010 | 117 | [42] |

[1] Google Scholar citations as of Sep. 5, 2019